\documentstyle[prl,aps,twocolumn,epsf]{revtex}

\begin{document}
\draft
\def\be{\begin{equation}}
\def\ee{\end{equation}}
\def\bea{\begin{eqnarray}}
\def\eea{\end{eqnarray}}

\def\et{ {\it et al.}}
\def\la{ \langle}
\def\ra{ \rangle}

\def\mdot{\ifmmode \dot M \else $\dot M$\fi}    
\def\mxd{\ifmmode \dot {M}_{x} \else $\dot {M}_{x}$\fi}    
\def\med{\ifmmode \dot {M}_{Edd} \else $\dot {M}_{Edd}$\fi}    
\def\bff{\ifmmode B_{f} \else $B_{f}$\fi}    

\def\apj{\ifmmode Astrophys. J. \else Astrophys. J.\fi}    
\def\apjl{\ifmmode Astrophys. J. Lett.
 \else Astrophys. J. Lett.\fi}    %
\def\aa{\ifmmode Astron. and Astrophys.
 \else Astron. and Astrophys.\fi}    %
\def\aaps{\ifmmode Astron. and Astrophys. Suppl.
 \else Astron. and Astrophys. Suppl.\fi}    %
\def\nat{\ifmmode Nature (London)\else Nature (London)\fi}
\def\prl{\ifmmode Phys. Rev. Lett. \else Phys. Rev. Lett.\fi}   
\def\prd{\ifmmode Phys. Rev. D. \else Phys. Rev. D.\fi}

\def\ms{\ifmmode M_{\odot} \else $M_{\odot}$\fi}    
\def\no{\ifmmode \nu_{1} \else $\nu_{1}$\fi}    
\def\nt{\ifmmode \nu_{2} \else $\nu_{2}$\fi}    
\def\ntmax{\ifmmode \nu_{2max} \else $\nu_{2max}$\fi}    
\def\nomax{\ifmmode \nu_{1max} \else $\nu_{1max}$\fi}    
\def\nh{\ifmmode \nu_{H} \else $\nu_{H}$\fi}    
\def\nz{\ifmmode \nu_{o} \else $\nu_{o}$\fi}    
\def\nht{\ifmmode \nu_{H2} \else $\nu_{H2}$\fi}    
\def\ns{\ifmmode \nu_{s} \else $\nu_{s}$\fi}    
\def\nb{\ifmmode \nu_{burst} \else $\nu_{burst}$\fi}    
\def\nkm{\ifmmode \nu_{km} \else $\nu_{km}$\fi}    
\def\dn{\ifmmode \Delta\nu \else $\Delta\nu$\fi}    
\def\rs{\ifmmode R_{s} \else $R_{s}$\fi}    
\def\rmm{\ifmmode R_{co} \else $R_{co}$\fi}    
\def\rim{\ifmmode R_{I} \else $R_{I}$\fi}    
\def\at{\ifmmode \alpha_{2} \else $\alpha_{2}$\fi}    
\def\ao{\ifmmode \alpha_{1} \else $\alpha_{1}$\fi}    
\def\ah{\ifmmode \alpha_{h} \else $\alpha_{h}$\fi}    
\def\akm{\ifmmode \alpha_{km} \else $\alpha_{km}$\fi}    
\def\kt{\ifmmode k_{2} \else $k_{2}$\fi}    
\def\ko{\ifmmode k_{1} \else $k_{1}$\fi}    

\def\gth{\ifmmode \gamma_{2H} \else $\gamma_{2H}$\fi}    
\def\gthmax{\ifmmode \gamma_{2Hmax} 
\else $\gamma_{2Hmax}$\fi}    

\def\c{\cite}

\title{ Quasi Periodic Oscillations in  X-ray Neutron Star \\  
and Probes of Perihelion Precession of General Relativity }
 
\author{ C. M. Zhang}
\vskip 0.5cm
\address{Instituto de F\'{\i}sica Te\'orica\\
Universidade Estadual Paulista\\
Rua Pamplona 145\\
01405-900\, S\~ao Paulo \\
Brazil\\
  zhangcm@ift.unesp.br}
\date{\today}
\maketitle

\begin{abstract}
We ascribe the twin kilohertz Quasi Periodic
Oscillations (kHz QPOs) of X-ray spectra of  Low Mass X-Ray Binaries
 (LMXBs) to the pseudo-Newtonian Keplerian frequency and the apogee and
perigee 
precession frequency of the same matter in the inner disk, and  ascribe
15 - 60 Hz QPO (HBO)  to the apogee (or perigee) 
precession and its second harmonic frequency to 
both apogee and perigee precession in the outer disk boundary of 
the neutron star (NS) magnetosphere. The radii of the inner  and outer 
disks are correlated each other  by  a factor of two is assumed. 
 The obtained conclusions include: all QPO frequencies increase  and 
frequency difference  of twin kHz QPOs decreases with increasing the
accretion rate. The obtained theoretical relations between HBO frequency 
and twin kHz QPOs are  simlilar to the measured  
empirical formula. Further, the theoretical formula to calculate  
 the NS mass by the twin kHz QPOs is proposed, and the resultant 
values are in the range of  1.4 to 1.8 $\ms$.
QPOs from LMXBs likely provide an accurate laboratory for a strong 
gravitational field, by which a new method to determine the 
 NS masses of LMXBs is suggested.  
\end{abstract}

\narrowtext

\par

Until now about twenty LMXBs have been discoveried
 to exhibit kHz QPOs, include one X-ray pulsar, with
  the Rossi X-ray Timing Explorer (RXTE) since the early
1996, briefly after its launch \cite{klis98,brad,wij99}.
The Z sources (Atoll sources), which are high (less)
luminous neutron-star   
low-mass X-ray binaries \cite{hk89}, typically show four
distinct types of QPOs \cite{klis98}. At present,
these are the  normal branch
oscillation (NBO) $\simeq 5-20$~Hz, the
horizontal 
branch oscillation (HBO) $\nh \simeq 15-60$~Hz 
\cite{klis98}, and the 
kHz QPOs $\nt(\no) \simeq 200-1200$~Hz 
that typically occur in  
pairs, upper frequency $\nt$ and lower frequency $\no$. 
In several Atoll sources, nearly 
coherent $\nb \simeq 330-590$~Hz oscillations have also
been detected during thermonuclear Type~I X-ray bursts, 
which are considered as the spin frequency of NS 
or twice of them.
 All of these QPOs but the burst oscillation have
centroid frequencies that increase with inferred mass
accretion rate \mdot, and are tightly correlated with each other 
\c{psa9902,psa9903}. 
In some cases, these correlations appear to depend
weakly on the other properties of the sources \c{psa9902,psa9903}.   
For example, the frequencies  $\nt$ and $\no$,
 as well as the frequencies  $\nt$ and $\nh$ 
 follow very
similar relations in five Z sources \c{psa9902,psa9903}. However
the frequency separation  
between the upper and the lower kHz QPOs $\dn \equiv \nt - \no$ 
decreases systematically 
with instantaneous \mdot{} in some cases, e.g. 
Sco X-1 \c{klis98,klis97}, 4U1608$-$52 and
4U1728$-$34\c{klis98,men99,men98}, 
then in the latter the observed 
coherent burst frequency 364 Hz is higher than its  maximum 
$\dn \sim $ 355 Hz \c{klis98,men98}.

A number of theoretical models have been
proposed to account for the  QPO phenomena  in 
X-ray NS systems.
For  the high frequency of kHz QPOs and its 
proportional relation to the accretion rate, simply, the 
upper kHz QPO $\nt$ is considered to originate
 from the Keplerian orbital frequencies at the preferred  radius   
close to the compact object, which exhibit 
the inner accretion flows, however the lower kHz QPO 
 $\no$ is  attributed to 
the beat of such frequency with the stellar spin $\ns$ 
 \c{mlp98}. 
 Recently, the  general
relativistic effects are paid much attetion 
to account for kHz QPOs \c{stel98,stel99}, which 
can explain the varied kHz QPOs separation $\dn$. 
Then the varied $\dn$  seems to exclude the simple beat
explanation for kHz QPOs \c{klis98,psa9903}, which predicts 
a  constant frequency separation.  
 However, it is also a common phenomenon 
for the  separation  
$\Delta\nu$ to be approximately constant.  
The range of $\Delta\nu$ is also 
quite narrow across different sources 
($\Delta\nu \approx 250-360$~Hz) with 
 a nearly coherent frequency of 
 $\nb \approx 330 - 590$~Hz. In some  cases 
(such as 4U 1702-43 and KS 1731-260) the burst frequency is
consistent, to within the errors, with the frequency separation 
$\Delta\nu$, or twice its value $2\Delta\nu$. So 
the approximately 
constant separation in some sources  seems also to make
the purely GR precession model in difficulty. 
Although many other viable new ideas are also proposed 
\c{tit98,others},
there has not yet been  one model satisfactorily  to explain
 all observational QPO  phenomena of LMXBs until 
now. The mechanisms of kHz QPOs of LMXBs are still  debated and open
problems.

Nonetheless, HBO frequency ($\nh \simeq 15 - 60 $Hz), first
discovered in GX 5-1 in 1985,
 is interpreted to be the beat frequency between
the Keperian frequency of the disk  and the stellar
spin frequency by the standard beat frequency model
(BFM)\c{alp85}.

It is commonly accepted that the kHz QPOs
 produce close to the innermost stable
orbit (IMS) or the surface of the NS, which will provide a probe to detect
the accretion flow around the non-Newtonian strong gravity region.

Here we concentrate on the explanation of kHz QPOs and HBOs (for
Atoll sources,  15 - 60 Hz QPO is supposed to be
the same mechanism as HBO of  Z sources\c{psa9902}), and neglect
the detail of the physical mechanism for QPO production.
 We ascribe $\nt$ to the
so-called pseudo-Newtonian Keperian frequency
proposed by Paczy$\acute{{\rm n}}$sky and  Wiita\c{pw,lai98}
$\nt \equiv \nu_K = (GM/4\pi^2 r^3)^{1/2}(1-2GM/rc^2 )^{-1}$, 
which is usually  applied in the black hole accretion
disk and
coalecing of compact objects\c{lai98} to mimic the general relativistic
effect,  and $\no$
 to the apogee + perigee precession frequency 
(${6\pi GM \over r c^{2}}$ rad per revolution) depicted
by the post-Newtonian scheme\c{will} $\no={6GM\over rc^{2} }\nt$. 
We suppose that both $\nt$ and $\no$
originate from the same matter in the inner disk boundary. Further,
contrary to the standard description to account  HBO for
BFM\c{alp85}, as well
as for the nodal precession\c{stel98} of Lense-Thirring effect 
in the inner disk, we
 assume that $\nh$  is an  apogee (or a perigee) precession frequency  
in the outer disk boundary (magnetosphere boundary).
 Then, on the reasons for deliminatimg BFM to account for HBO,  we
have the following arguements. First,
the simple BFM cannot predict the second harmonic frequency $\nht$,
which was detected in Sco X-1 with 90 Hz corresponding to 
HBO 45 Hz\c{klis97}.
Second,
$\nh$, $\nt$ and $\no$ correlations seems to weakly depend on the 
stellar spin frequency \c{klis98,psa9902}. Third, BFM should predict the
null or extremely 
low $\nh$ if the accretion rate is extremely low, where the inverse 
of the spin-up
torque should arise $\nh$ to be inversely correlated to the 
accretion rate. But this phenomenon has never been 
detected in both Z and Atoll 
sources, and the lowest detected HBO $\nh \sim 15 $Hz with the lower  
accretion rate hinted in color-color diagram of X-ray spectra\c{hk89}. 
Well, here  the proposed  periastron precession model (PPM) can ascribe
the 
second harmonic $\nht$ to  the precession frequency of both apogee and
perigee at the outer disk boundary, which is twice of HBO ($\nht=2\nh$).  

If this PPM interpretation were 
confirmed,  QPOs would provide a probe into the  test for
strong field  GR, by which we can determine or constrain the NS
parameters,
such as mass, radius and magnetosphere radius, as well as the magnetic
field strength.  We set $c=G=1$ throughout  this paper.

\par

On account of the complexity of the accretion flow close to the 
innermost stable orbit $\rim = 3\rs$, where $\rs = 2GM/c^2 = 3m$ (km)
 is the Schwarzschild radius calculated with  the gravitational mass 
 M (m is the mass in unit of solar mass), 
the motion of the disk matter will be  
influenced by not only gravitaional field but the 
star magnetic field, which should be complicated. 
For simplicity, we
assume that 
the disk matter to exhibit QPO are mainly dominated 
 by the Schwarzschild  gravitational field with 
 a slightly eccentric orbit ($e\simeq 0$). Therefore, 
 these QPO frequencies
are conveniently arranged as follows with the consideration of 
the possible modification.
\be
$$\no = 3\times \ao{\; } {y} \nt - \ko \nh\;, 
$$
\label{no}
\ee
\be
$$\nt = \at{\;}{\nz}  y^{3/2}(1 - y)^{-1} - \kt \nh\;,
$$
\label{nt}
\ee
\be
$$\nh = {3\over 2}\times \ah{\;} {(\phi y)} \nkm \;,\;\;\; $$
$$\phi \equiv {r \over \rmm}
$$
\label{nh}
\ee
\be
$$\nkm = \akm {\;} {\nz}(\phi y)^{3/2}(1 - \phi y)^{-1}\;,
$$
\label{nkm}
\ee
\be
 $$y\equiv {R_{s} \over r} \mxd^{2/7}\;,\;\;\; $$
$$\nz \equiv {11300\over m}{\;}{(Hz)}\;,$$
\ee
where r is the inner disk radius, and $\rmm$ is the   outer
disk radius where it is corotating with the  magnetosphere.
$\nkm$ is the pseudo-Newtonian Keplerian frequency of the outer
disk. 
$\ao, \at, \ah, \akm \sim 1$ are the modification coefficients on 
account of the orbit eccentricity and other approximated 
uncertainty, which might be 
 almost unity under the ideal condition with the  circular orbit. 
$\ko(\kt) = 0, 1, 2 $ represent the possible beat 
modes between kHz QPOs and HBO.    
 $\phi$ is a scaling parameter to connect the inner and outer
disks, but here we suppose it to be 0.5 for the reason of the best
fitting.
 $y$ is the ratio of the Schwarzshild radius to the instantaneous 
inner disk radius, wich is proportionally related to the 
fractal accretion rate, $\mxd\equiv {\mdot \over \la\mdot\ra}$,  
the ratio between the instantaneous accretion rate $\mdot$ and the average 
accretion rate $\la\mdot\ra$, and the latter determines the  
corotation magnetosphere radius 
 $\rmm = 1.9\times 10^{6} (cm) B_{8}^{4/7}
\la\mdot\ra^{-2/7}m^{1/7}R_{6}^{12/7}$ (with the nearly unity 
fastness\c{gl79,st}) 
, where $B_{8}$ and  $R_{6}$ are star magnetic field and radius in unit 
of $10^8$ G and $10^6$ cm respectively. 

The relations $\nh$ vs.  $\nt$  and $\dn$ vs. $\no$ are plotted in 
Fig.1 and Fig.2, together with the well measured  source samples, 
and it is shown that the agreement  bewteen the model and 
the observed QPO
data is quite well for the   selected values of the 
NS mass from 1.4 to 1.8 $\ms$, which is the  
solely free parameter in the Eqs.(1-5). 
 From Eqs.(\ref{no}), (\ref{nt}) and (\ref{nh}), we can derive the
thoeretical
relations between QPO frequencies   in the
following if we set the ideal parameter conditions with
$\ao=\at=\ah=\akm= 1$ and 
$\ko=\kt=0$. 
\be
$$\nh \simeq 44.2{\;}(Hz){\;}({\no\over 500})
[ 1 - 0.1 ({m\no \over 500})^{2/5} ] \;,
$$
\label{nhno}
\ee
\be
$$\nh \simeq 51.5{\;}(Hz){\;}m^{2/3}({\nt\over 1000})^{5/3}
[1 - {1\over 3}({m\nt\over 1000})^{2/3}]^{2/3} \;,
$$
\label{nhnt}
\ee
\be
$$\no \simeq 595{\;}(Hz){\;}m^{2/3}({\nt\over 1000})^{5/3}
[1 - {1\over 6}({m\nt\over 1000})^{2/3}]^{2/3} \;.
$$
\label{nont}
\ee
Further, we can  also obtain the NS mass formula represented  by 
the QPO frequencies without consideration of the other 
modification parameters 
\be
$$m = 2.17 \times({\no\over \nt})^{3/2}({\nt\over 1000})^{-1}
[1 - {\no\over 3\nt}]^{-1} \;,
$$
\label{mot}
\ee
\be
$$m \simeq 1.56 \times({\nh\over 70})^{3/2}({\nt\over 1000})^{-5/2}
[1 - {6\sqrt{2}\nh\over \nt}]^{-1} \;,
$$
\label{mht}
\ee
For Z sources (almost Eddington accretion), $\mxd =1 $ corresponds to 
the maximum
QPO frequencies, then  we can determine $y_{max}$ by the observed $\ntmax$ 
and calculate the corotation magnetosphere radius, as well as the spin
frequency. It is stressed that the observed maximum upper kHz QPOs are  
most likely the  maximum kHz QPO frequencies for Z sources but they are 
 not
neccessary. However  for Atoll sources (lower accretion rate), 
they have  no
above
properties shared by  Z sources. But, we can apply the measured $\ntmax$ 
to predict  the possibly upper limit for the burst frequency (spin) 
of Atoll sources. Moreover, 
 for the  known burst frequency (spin), the corotation  
Keplerian frequency corresponds to $\mxd=1$, by which we can determine the 
magnetosphere radius and magnetic field strength of the source.

The estimation of the NS magnetic field strength\c{zcm98} 
 is given by  $B = ({\rmm\over R})^{7/4}\bff$ with 
$\bff= 4.3\times10^{8}\;
(G)\;(\la\mdot\ra/\med)^{1/2}m^{1/4}R_{6}^{-5/4}$, 
where R is star radius.
The calculated 
NS parameters of LMXBs  are list in TABLE I for six Z sources and eleven 
Atoll sources. 
 The obtained magnetic fields for both Z and Atoll sources
 are about $2\times 10^9$ G
 and $2\times 10^8$ G respectively, 
which are  consistent with the originally 
suggested values\c{hk89,wz97}. 
It is also interested that the derived spin
frequencies are in the range of 289 (Hz) - 356 (Hz) with hinted
magnetosphere radius in the range of 37 (km) - 47 (km), and the
homogenous span of spins of LMXBs is consistent with the period
distribution of millisecond pulsars. However the reason for this
homogeneity is still in studying\c{wz97}.

Further, we can define the maximum kHz QPO frequency 
($\ko=\kt=0$ mode) at the innermost 
stable orbit $\rim=3\rs$ or y=1/3, then  
$\ntmax=\nomax=\nt(y=1/3)=1625\times(2/m){ }(Hz)$, which is much
higher than the measured maximum kHz QPO frequency 1228 Hz
(4U1636-53).  The higher $\ntmax$ seems to hint the saturation of 
$\nt$ will not happen below 1200 Hz if the levelling off of the 
X-ray signal 
orginates from the IMS. The recent observation on the 
kHz QPO saturation obtained the debated conclusions\c{men98,zhan98}.

It is easy to see in FIG.2 that the kHz QPO separation 
$\Delta\nu$ vs. $\no$  for two well measured sources 
Sco~X-1 and 4U1728-34\c{klis98,klis97,men99} is plotted, 
together with the proposed PPM theoretical curves. 
It is apparent that for NS masses at about 1.45$\ms$ and 1.70$\ms$,
 the simple model outlined above 
is in qualitative agreement with the measured 
values, including the  decrease of $\Delta\nu$ for increasing $\no$.
 For the  separation  
$\Delta\nu$ to be approximately constant, we argue that the beat mode
with $\ko=1$ and $\kt=0$ can mimic the almost constant
separation  in  the range of $\no= 500 \sim 900 (Hz)$, 
corresponding to the
parameter range of 
$y= 0.2 \sim 0.25 $. For the mode $(\ko=1,\;\kt=0)$, the separation 
becomes $\dn(y)\simeq (1-2.75y)\no(\ko=0)$, and we will find 
${\dn(y=0.2)/\dn(y=0.25)} \simeq 1$. If the spin frequency is defined 
by the corotation frequency of magnetosphere-disk of about 40 (km) with
 y=0.25, 
i.e. $\ns=\nkm(y=0.25)$, then we obtain that the ratios  
${\dn(y=0.2)/\ns} \simeq 1.01$ and ${\dn(y=0.25)/\ns} \simeq
0.99$, both ratios are independent of the NS mass m ! This may be the
reason why 
for a few  sources the measured $\dn$ are similar to the burst
frequencies within the errors. This coincidence reflects the homogenous 
magnetospheres of LMXBs in the range of about 40 (km).   
\vskip 0.3cm

{TABLE I. The parameters of six  Z sources \\and eleven  Atoll sources}
\vskip 0.1cm
\begin{tabular}{|c|ccccc|}\hline
Source & $\ntmax^{*}$ & m & $\rmm$ &$\ns $ & $B_{8}^{+}$\\
       & (Hz)      &($\ms$)& (km)   & (Hz)  & (G)\\\hline
GX~5$-$1   & 930$^{a}$ &  1.71 &   46.7  & 289  & $ 22$\\\hline
GX~340$+$0 & 945$^{a}$ &  1.43 &   42.5  & 297  & $ 18$\\\hline
Cyg~X-2    & 1005  &  1.50 & 42.1  & 313  & $ 17$\\\hline
GX~349$+$2 & 1020  &  1.78 & 45.6  & 314  & $ 21$\\\hline
GX~17$+$2  & 1085  &  1.63 & 42.3  & 334  & $ 18$\\\hline
Sco~X-1    & 1130  &  1.75 & 43.0  & 345  & $ 19$\\\hline\hline
4U0614$+$09 & 1145 &  1.41 &   38.2  & 356$^{\dag}$  & $ 1.5$\\\hline
4U1608$-$52 & 1090 &  1.65 &   42.4  & 335$^{\dag}$  & $ 1.8$\\\hline
4U1705$-$44 & 1075 &  1.61 &   42.2  & 332$^{\dag}$  & $ 1.8$\\\hline
4U1735$-$44 & 1150 &  1.62 &   41.0  & 352$^{\dag}$  & $ 1.7$\\\hline
4U1820$-$30 & 1065 &  1.55 &   41.6  & 330$^{\dag}$  & $ 1.7$\\\hline
4U1915$-$05 & 1005 &  1.37 &   40.3  & 316$^{\dag}$  & $ 1.6$\\\hline
XTEJ2123$-$058 & 1140 &  1.70 &42.2  & 349$^{\dag}$  & $ 1.8$\\\hline
4U1636$-$53   && 1.75   &   45.5  & 290$^{*}$ & $ 2.1$\\\hline
4U1702$-$43   && 1.49   &   39.4  & 330$^{*}  $ & $ 1.6$\\\hline
4U1728$-$34   && 1.44   &   36.9  & 364$^{*}  $ & $ 1.4$\\\hline
4U1731$-$260  && 1.50   &   45.5  & 524/2$^{*}$ & $ 2.0$\\\hline
\end{tabular}\\
\vskip 0.1cm
{a }:{ } Inferred from the frequency  separation of kHz QPOs. 
{*}: { }data from van der Klis  
\c{klis98}, and references therein.
$\dag$ :{ } the possibly upper limit for the burst frequencies(or half
of them).{ } {+}:{ } star radii are  assumed to be 
15 km with  the Eddington accretion rate $\med$ for Z sources and 
0.01$\med$ for Atoll sources. 
\vskip 0.3cm

\par

In summary, we state that the model described here is  simply and
roughly one, and many physical
details are neglected, such as NS spin induced gravitomagnetic effect, 
NS quadrupole induced nodal precession\c{stel98}, the self-gravity of the disk, 
magnetosphere structure and magnetic axis inclination, the spiral-in effect 
of the accreted matter, the origination and influence of the non-zero
eccentricity, etc.. 
Especially, the non-zero eccentricity should have somewhat effects on the 
QPOs, but the origination mechanism for this is still unknown. At least,
we can
speculate that the motion of the accreted matter in the disks 
might be  not exactly described by  
a free test particle in a circular orbit of a purely gravitational field.   
Consideration of these factors will construct our future exploration 
and understanding  of QPOs in LMXBs. 

Finally we stress  that the proposed PPM  model has also 
implications  that the
 QPO phenomenon in LMXBs is likely to reflect the 
fundamental  relativistic motion of matter 
in the vicinity of the NS strong gravitational field. 
As discussed in
 this  paper, 
if the PPM were successfully to explain  the observed   QPOs in LMXBs, we
not 
only for the second time  uncover the GR periastron precession evidence 
in the NS besides that in the Mercury orbit of the solar system,
predicted by Einstein, but also, 
by means of QPO, we developed  a new method to determine the NS
mass from Eq.(\ref{mot}) and Eq.(\ref{mht}), as a contrast  
to determining NS mass in NS binary system such as PSR 1913+16. 
\vskip .3cm

The author  is  grateful to E. Ford,  M. M\'endez,  R. Wijnands and 
W. Zhang for critical reading of the manuscript and helpful discussions.

\begin{figure}
\epsfxsize=3.in
\vskip -7.cm
\epsffile{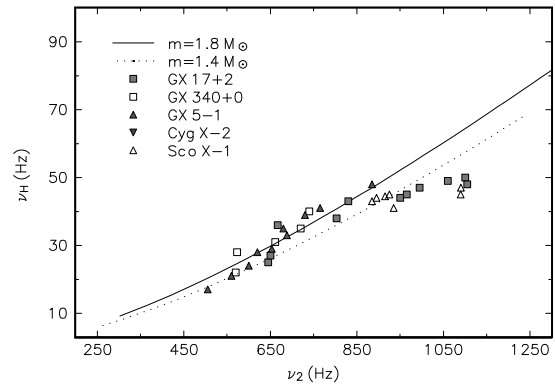}
\caption[fig1]{ HBO frequency $\nu_H$ versus the upper kHz QPO 
frequency $\nt$ for five Z sources of  LMXBs. 
(cf.[4,5]  and references therein). 
Error bars are not plotted for the sake of clarity.
The model presents a well fitting for  the  nearly circular 
orbit of NS mass  from about 1.4  to 1.8 solar mass.}
\label{fig1}
\end{figure}

\begin{figure}
\epsfxsize=3.in
\vskip -6.5cm
\epsffile{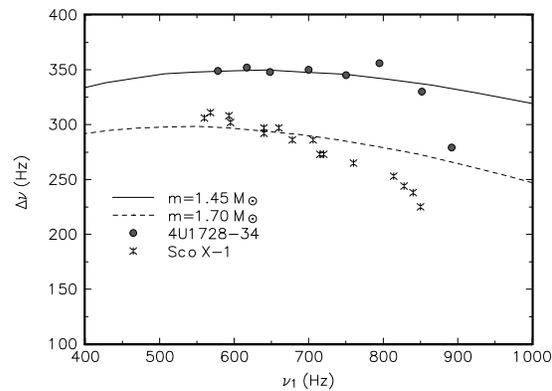}
\caption{ kHz QPO frequency separation $\dn$ versus the lower kHz QPO
frequency $\nt$ for Sco X-1 and 4U1728-34 (cf.[1,6-8]). 
Error bars are not plotted for the sake of clarity.}
\label{fig2}
\end{figure}
\end{document}